\documentclass[aps,prd,preprint]{revtex4}
\usepackage{epsfig}

\usepackage{graphicx,psfrag}
\usepackage{amsmath}
\usepackage{bm}

\newcommand{\beqa}{\begin{eqnarray}}
\newcommand{\eeqa}{\end{eqnarray}}
\newcommand{\beq}{\begin{equation}}
\newcommand{\eeq}{\end{equation}}
\newcommand{\bsp}{\begin{split}}
\newcommand{\esp}{\end{split}}
\newcommand{\bal}{\begin{align}}
\newcommand{\eal}{\end{align}}

\begin{document}

\title{$1/N_c$ expansion and the spin-flavor structure of the quark interaction in the constituent quark model}

\def\addDP{National Institute for Physics and Nuclear Engineering, 
Department of Particle Physics, 077125 Bucharest, Romania}
\def\addCS{CONICET - Departamento de F\'{\i}sica, FCEyN, Universidad de Buenos Aires, 
Ciudad Universitaria, Pab.~1, (1428) Buenos Aires, Argentina}

\author{Dan Pirjol}\affiliation{\addDP}
\author{Carlos Schat}
\affiliation{Department of Physics and Astronomy, Ohio University, Athens, Ohio 45701, USA}
\affiliation{\addCS}

\begin{abstract} \vspace*{18pt}
We study the hierarchy of the coefficients in the $1/N_c$ expansion for the negative parity $L=1$
excited baryons from the perspective of the constituent quark model. This is related to the
problem of determining the spin-flavor structure of the 
quark interaction.
The most general two-body scalar interaction between quarks contains the spin-flavor structures
$t_1^a t_2^a, \vec s_1\cdot \vec s_2$ and $ \vec s_1\cdot \vec s_2 t_1^a t_2^a$.
We show that in the limit of a zero range interaction all these structures are matched onto
the same hadronic mass operator $S_c^2$, which gives a possible explanation for the dominance of 
this operator in the $1/N_c$ expansion for the $L=1$ states and implies that in this limit 
it is impossible to distinguish between 
these  different spin-flavor structures. Modeling a finite range interaction through the
exchange of a vector and pseudoscalar meson, we propose a test for the spin-flavor
dependence of the quark forces.
For the scalar part of the quark interaction we find that both pion exchange and gluon exchange
are compatible with data.
\end{abstract}
\maketitle

\section{Introduction}

The application of the $1/N_c$ expansion to the excited baryons sector has produced a number
of interesting results, see Refs.~\cite{Pirjol:2007ui} for a recent review. 
Baryon properties like masses or axial couplings can be expanded in a systematic 
way using an explicit representation of operators acting on quark degrees of freedom 
\cite{Carone:1993dz} \cite{Luty:1993fu} \cite{Dashen:1994qi} \cite{Goity:1996hk}. 
Working to order ${\cal O}(1/N_c)$, there are two  ${\cal O}(N_c^0)$  
and eight ${\cal O}(1/N_c)$ operators
in the expansion of the mass operator of the nonstrange $L=1$ excited baryons \cite{Carlson:1998vx}.
In this paper we will be concerned with the
observed pattern of the coefficients of the various operators in the $1/N_c$ expansion
when applied to the study of these negative parity excited states.

The most prominent feature of the coefficients $c_i$ is the dominance of the $O(1/N_c)$ 
operator $\frac{1}{N_c} S_c^2$, which is also confirmed by extending the analysis to 
flavor $SU(3)$, including all the members of the {\bf 70}-plet \cite{Schat:2001xr} \cite{Goity:2002pu}.
The coefficients of the ${\cal O}(N_c^0)$ and of the other ${\cal O}(1/N_c)$ operators are smaller than expected by $1/N_c$
power counting alone, as their natural size is set by the coefficient of the unit operator and is of the order 
of  $\sim 500$ MeV. The dominance of the $S_c^2$ operator has been explained in Ref.~\cite{Carlson:1998vx}
by assuming dominance of a pion-mediated interaction among constituent quarks. 

In this paper we propose another explanation for this hierarchy of the coefficients: the short range of 
the quark
interaction in the constituent quark model. We show that in the limit of a contact interaction any scalar
quark interaction, regardless of its spin-flavor structure, is matched onto the single operator $S_c^2$. 
This implies the surprising conclusion that, within the scalar interactions,
it is impossible to distinguish between quark interactions with different spin-flavor structures, 
such as the one-gluon exchange model (OGE) \cite{De Rujula:1975ge} \cite{Jaffe:1976ih}
and the Goldstone 
boson exchange model (GBE) \cite{Glozman:1995fu}, as long as these interactions are of very short range. 

On the other hand, a more complex spatial dependence of the quark forces will introduce two other operators $t_1^a T_c^a$ and 
$\vec s_1 \cdot \vec S_c$. Their strengths depend on the range of the interaction \cite{Schat:2001xr}  
and are sensitive to the spin-flavor structure. 
Modeling the quark interaction as
mediated by the exchange of a meson of mass $\mu$ we discuss the consequences of a finite range given by $1/\mu$
and propose
the sign of the ratio of two coefficients as a test for the spin-flavor structure of the interaction.
We finally use the wave functions of the Isgur-Karl model \cite{Isgur:1977ef} 
with  a harmonic oscillator potential 
to compute this ratio and constrain the mass scale $\mu$.

The paper is organized as follows. In Sec.~\ref{zeror} we discuss the matching of the three possible spin-flavor 
structures to the effective operator expansion and point out that in the case of a zero 
range interaction only one operator dominates. In Sec.~\ref{finiter} we discuss the finite range correction and propose
a test for the spin-flavor structure of the interaction.
In Sec.~\ref{IKcalc} we perform a model calculation of the orbital reduced matrix elements. 
In Sec.~\ref{concl}  we summarize and present our conclusions.

\begin{table}[b]
\begin{tabular}{cccc}
\hline \hline
 ${\cal O}_{ij}$ & $O_S$ & $O_{MS}$ &  \\
\hline
$t_i^a t_j^a$ & \qquad $\frac12 T^2 - \frac32 C_2(F)$ & $-T^2 + 3 t_1^a T_c^a + 3 C_2(F)$ 
       & $\frac12 O_1, \frac12 O_1 - \frac32 O_2$ \\
$\vec s_i \cdot \vec s_j$ & $\frac12 \vec S^2 - \frac98$ 
       & $-\vec S^2 + 3 \vec s_1 \cdot \vec S_c + \frac94$ 
       & $\frac12 O_2 + O_3, -O_2 + O_3$ \\
$\vec s_i \cdot \vec s_j t_i^a t_j^a$ \qquad & \qquad $\frac12 G^2 - \frac98 C_2(F)$ \qquad
       &\qquad $3g_1^{ka} G_c^{ka} - G^2 + \frac94 C_2(F)$ \qquad 
       & $-\frac18 O_1 - \frac{1}{4F}O_2 - \frac{1}{2F}O_3$ \,, \\
 & & & $-\frac18 O_1 + (\frac38 + \frac{1}{2F}) O_2 - \frac{1}{2F}O_3$ \\
\hline\hline
\end{tabular}
\label{TableI}
\caption{The projection of the most general scalar quark interaction onto irreducible representations of $S_3$
allows to express the corresponding reduced matrix elements as matrix elements of the operators listed in 
the second and third column. They are shown again in the last column written in terms of $O_1,O_2,O_3$ defined in the
text, up to terms proportional to the unit operator.
The quadratic Casimir of the fundamental representation of the flavor group $SU(F)$ is 
$C_2(F) = (F^2-1)/(2F)$.}
\end{table}

\section{Zero range scalar quark interactions}
\label{zeror}

The most general quark Hamiltonian containing only two-body interactions
has the form \cite{Pirjol:2008gd}
\begin{eqnarray}\label{1}
H_{qq} = H_0 + \sum_{i<j} (f_1(\vec r_{ij})  t_i^a t_j^a + 
f_2(\vec r_{ij})  \vec s_i \cdot \vec s_j + 
f_3(\vec r_{ij})  \vec s_i t_i^a \cdot \vec s_j t_j^a)  + H_{s-o} + H_q \, ,
\end{eqnarray}
where $\vec r_{ij}=\vec r_i-\vec r_j$ is the distance between quarks $i,j$
and  $H_0$ is the part of the quark Hamiltonian which does not depend on the quarks spin
and flavor degrees of freedom. We show explicitly only the part of the Hamiltonian
which transforms as a scalar ($\ell=0$) under $SO(3)$, the group of orbital rotations - the scalar part 
of the quark Hamiltonian. The $H_{s-o}, H_q $ denote the spin-orbit and the quadrupole interaction,  
which transform as a vector ($\ell=1$) and a traceless and symmetric 
tensor of rank two ($\ell=2$)  under $SO(3)$, 
respectively.

We will consider in this Section the case of a contact scalar interaction
\begin{eqnarray}
f_\nu (\vec r_{ij}) = A_\nu \delta^{(3)}(\vec r_{ij})\,,\qquad \nu=1,2,3,
\end{eqnarray}
and study the following question: what information can be obtained from the coefficients
$c_k$ of the $1/N_c$ studies of the spectrum of $L=1$ negative parity baryons? The motivation for this
investigation is related to the question of distinguishing between different models of the  
quark interaction. The two main models considered in the literature are: i) the one-gluon
exchange model (OGE)  \cite{De Rujula:1975ge}\cite{Jaffe:1976ih}
, and ii) the Goldstone-boson exchange model (GBE) \cite{Glozman:1995fu}. 
In this paper we will consider a wider class of models, corresponding to the
most general two-body interaction with arbitrary spin-flavor structure.

Our analysis will be completely
general, and will not make any assumptions about the orbital wave functions of these states.
We will use the method described in Ref.~\cite{Pirjol:2007ed} for obtaining predictions
in the quark model by exploiting the transformation properties of the states and interaction
Hamiltonian under $S_N$, the permutation group of the $N$ quarks. 
The application of the $S_3$ symmetry in this context was also considered in Ref.~\cite{Collins:1998ny}.
In particular, this
allows one to match any quark Hamiltonian onto the operators of the $1/N_c$ expansion. 
The mass operator in the $1/N_c$ expansion has also been compared with the predictions
of a particular quark model in Refs.~\cite{Semay:2007cv,Semay:2007ff} using a different approach.
We give in the following a brief summary of the results of Ref.~\cite{Pirjol:2007ed} that will be used in this work.

Consider a general two-body quark Hamiltonian of the form
\begin{eqnarray}
H_{qq} = \sum_{i<j} \sum_\nu f_\nu(\vec r_{ij}) O_{ij}^{(\nu)} \,,
\end{eqnarray}
where $O_{ij}^{(\nu)}$ act only on the spin-flavor degrees of freedom of the quarks $i,j$,
and $f_\nu(\vec r_{ij})$ act only on their orbital degrees of freedom. The index $\nu$
runs over all distinct spin-flavor structures.
Using the transformation properties of the states and operators under $S_N$,
the permutation group of $N$ objects, it has been shown in Ref.~\cite{Pirjol:2007ed}
that the mass operator corresponding to the Hamiltonian $H_{qq}$ has for $N_c=3$ the general form
\begin{eqnarray}
M = \frac13 \sum_\nu \left(  R_S^{(\nu)} O_S^{(\nu)} + R^{(\nu)}_{MS} O_{MS}^{(\nu)} \right) \,,
\end{eqnarray}
where $O_S^{(\nu)} ,O_{MS}^{(\nu)} $ ($R^{(\nu)}_S, R^{(\nu)}_{MS}$) are the 
reduced matrix elements of the projections of the spin-flavor operators $O_{ij}^{(\nu)}$
(orbital operators $f_\nu(\vec r_{ij})$) onto the $S,MS$ irreps of $S_N$. 
For an explicit example see Ref.~\cite{Galeta:2009pn}.

Table~\ref{TableI} lists all possible scalar two-body spin-flavor 
operators $O_{ij}$ and their projections onto irreducible representations of $S_3$.
The projections can be all expressed in terms of the 
three operators
\begin{eqnarray}\label{Opsi}
O_1 &=& T^2 \,,\qquad
O_2 = S_c^2 \,,\qquad
O_3 = \vec s_1 \cdot \vec S_c \,,
\end{eqnarray}
such that the quark Hamiltonian $H_{qq}$ is matched onto the hadronic mass
operator
\begin{eqnarray}\label{scalar}
M = c_0 \mathbf{1} + c_1T^2 + c_2 S_c^2 + c_3 \vec s_1 \cdot \vec S_c + \cdots
\end{eqnarray}
The ellipses denote terms arising from the tensor and spin-orbit interactions, which 
are not considered here \footnote{A detailed study of the spin-orbit interaction was presented in
Ref.~\cite{Pirjol:2008gd}. In this paper we will focus exclusively on the scalar interaction.
}. 
The operators $O_i$ in Eq.~(\ref{Opsi}) have been introduced in the context of the
$1/N_c$ expansion for the negative $L=1$ baryons in Ref.~\cite{Carlson:1998vx},
where the matrix elements of these operators on the relevant states have been
computed.
Although we use the notation of this paper, in the present discussion we will have
$N_c=3$ throughout.

The reduced matrix elements of the scalar orbital operators are defined in terms of the matrix
elements of $f_\nu(\vec r_{12})$ taken between a basis of orbital wave functions $\chi_{2,3}$
transforming in the $MS$ irreducible representation of $S_3$
\begin{eqnarray}\label{RSMSdef}
\langle \chi_i |f_\nu(\vec r_{12}) |\chi_j \rangle = \frac13
\left(
\begin{array}{cc}
2(R^{(\nu)}_S + R^{(\nu)}_{MS}) & R^{(\nu)}_S + R^{(\nu)}_{MS} \\
R^{(\nu)}_S + R^{(\nu)}_{MS} & 2R^{(\nu)}_S - R^{(\nu)}_{MS} \\
\end{array}
\right) \, .
\end{eqnarray}
The basis $\chi_{2,3}$ is defined by its transformation properties under $S_3$ given 
in Eqs.~(6)-(8) of Ref.~\cite{Pirjol:2007ed}, and is normalized according to $\langle \chi_i | \chi_j \rangle = 1 + \delta_{ij}$.

The coefficients of the operators appearing in the scalar part of the mass operator
Eq.~(\ref{scalar}) are
\begin{eqnarray}\label{ciscalar}
c_1 &=& \frac16 (R^{(\nu)}_S + R^{(\nu)}_{MS}) \left\{
\begin{array}{c}
1 \\
0 \\
-\frac14 \\
\end{array}
\right\}_\nu \,,\qquad
c_3 = \frac16 (R^{(\nu)}_S + R^{(\nu)}_{MS}) \left\{
\begin{array}{c}
0 \\
2 \\
-\frac{1}{F} \\
\end{array}
\right\}_\nu\,,\\
c_2 &=& \frac16 (R^{(\nu)}_S + R^{(\nu)}_{MS}) \left\{
\begin{array}{c}
-\frac32 \\
-\frac12 \\
\frac38 + \frac{1}{4F} \\
\end{array}
\right\}_\nu + \frac16 (R^{(\nu)}_S - R^{(\nu)}_{MS}) \left\{
\begin{array}{c}
\frac32 \\
\frac32 \\
-\frac38 - \frac{3}{4F} \\
\end{array}
\right\}_\nu \,,
\label{c2scalar}
\end{eqnarray}
where the index $\nu=1,2,3$ corresponds to the three possible 2-body operators $O_{ij}^{(\nu)} = t_i^a t_j^a,
\vec s_i \cdot \vec s_j, \vec s_i t_i^a \cdot \vec s_j t_j^a$.

Taking the index $\nu=2$ corresponds to the OGE model, and $\nu=3$ to the GBE model. We note the
following relations for the coefficients $c_i$, already pointed out in Ref.~\cite{Pirjol:2008gd},
which hold irrespective of the orbital dependence of the interactions
\begin{eqnarray}
\mbox{OGE} : && c_1 = 0 \,, \\
\mbox{GBE} : && c_1 = \frac{F}{4} c_3 \,.
\end{eqnarray}

The numerical values of the reduced matrix elements $R_S, R_{MS}$ depend on the
detailed form of the hadronic wave functions, and of the spatial functions $f(\vec r_{ij})$.
It has been shown in Ref.~\cite{Galeta:2009pn} that in the case of a contact interaction
$ f(\vec r_{ij}) \sim \delta^{(3)}(\vec r_{ij})$, the symmetric and mixed symmetric reduced matrix elements
$R_S, R_{MS}$ are related as
\begin{eqnarray}\label{RSMS}
R_S = - R_{MS} \,.
\end{eqnarray}
We recall briefly the proof of this relation, which follows from the formula $\langle \chi_2 |
f(\vec r_{12}) | \chi_2 \rangle = \frac23 (R_S + R_{MS})$, see Eq.~(\ref{RSMSdef}). The basis of
MS states $\chi_{2,3}$ is defined such that $P_{12} \chi_2 = - \chi_2$, which implies that
$\chi_2$ is antisymmetric under an exchange of the quarks 1,2, and thus it vanishes for $\vec r_{12}=0$.
This implies that for a contact interaction $f(\vec r_{12}) \sim  \delta^{(3)}(\vec r_{12})$, the relation Eq.~(\ref{RSMS}) 
holds among the two reduced matrix elements $R_S, R_{MS}$.

Using the relation Eq.~(\ref{RSMS}) we find that the coefficients $c_{1,2,3}$ are given, in the limit of a 
contact scalar interaction, by
\begin{eqnarray}
c_1 = c_3 = 0\,,\qquad
c_2 =  \frac13 R^{(\nu)}_S \left\{
\begin{array}{c}
\frac32 \\
\frac32 \\
-\frac38 - \frac{3}{4F} \\
\end{array}
\right\}_\nu \,.
\end{eqnarray}

Very surprisingly, all three possible zero range two-body interactions  $O_{ij}^{(\nu)} = t_i^a t_j^a,
\vec s_i \cdot \vec s_j, \vec s_i t_i^a \cdot \vec s_j t_j^a$ are matched onto the
same operator $O_2 = S_c^2$ in the effective theory! This means that there
is  no way to distinguish between these three types of scalar interactions if they are
contact interactions. 

Experimentally, at $N_c=3$ one can determine
only two linear combinations of the three coefficients $c_{1,2,3}$ (as functions of
$\theta_{N1}$) \cite{Pirjol:2008gd} from the mass spectrum and 
mixing angles of the negative parity $L=1$ baryons, which can be taken as
\begin{eqnarray}
\tilde c_1 &=& c_1 - \frac12 c_3 \,,\qquad
\tilde c_2 = c_2 + c_3\,.
\end{eqnarray}
This choice corresponds to eliminate the operator
$O_3 = \vec s_1 \cdot \vec S_c$ using the exact relation 
$T^2 - 2 S_c^2 + 2 \vec s_1 \cdot \vec S_c =  \delta c_0 \mathbf{1} $ with $\delta c_0 = -\frac94$ 
for $F=2 $ and $\delta c_0 = 0$ for $F=3$, 
that holds on the physical states 
\footnote{The general $SU(2)$ expression for arbitrary $N_c$ 
is $T^2 - \frac{N+1}{N-1} S_c^2 + 2 \vec s_1 \cdot \vec S_c = - \frac{N+15}{4(N-1)} \mathbf{1} $ and holds only on 
the states that are identified with the physical states when $N_c=3$. }.

The coefficients $\tilde c_{1,2}$ can be expressed in terms of the nonstrange hadron masses
and mixing angles as
\begin{eqnarray}\label{c1tildeexp}
\tilde c_1 &=& \frac{1}{18}( - N(1535) \sin^2\theta_{N1} - N(1650) \cos^2 \theta_{N1}
- 2 N(1520) \sin^2\theta_{N3} \\
&-& 2 N(1700) \cos^2\theta_{N3} - 3N_{5/2} + 2 \Delta_{1/2} + 4 \Delta_{3/2}) \,, \nonumber \\
\label{c2tildeexp}
\tilde c_2 &=& \frac{1}{6}( (N(1535) - 2 N(1650)) \sin^2\theta_{N1} + (N(1650) - 2N(1535)) \cos^2\theta_{N1} \\
&+& (2N(1520) - 4N(1700)) \sin^2\theta_{N3} + (2N(1700) - 4 N(1520)) \cos^2\theta_{N3} +
3 N_{5/2}) \,. \nonumber
\end{eqnarray}
The mixing angles $\theta_{N1, N3}$ are related by the correlation 
\begin{eqnarray}\label{correlation}
&& \frac12 (N(1535)+N(1650)) + \frac12 (N(1535) - N(1650)) (3\cos 2\theta_{N1} + \sin 2\theta_{N1}) \\
&& - \frac75 (N(1520) + N(1700)) + (N(1520) - N(1700)) (-\frac35 \cos 2\theta_{N3} + 
\sqrt{\frac52} \sin 2\theta_{N3} ) \nonumber \\
&& = - 2\Delta_{1/2} + 2\Delta_{3/2} - \frac95 N_{5/2}\,.\nonumber
\end{eqnarray}
Eqs.(\ref{c1tildeexp},\ref{c2tildeexp},\ref{correlation}) 
 hold in the most general constituent quark model containing only two-body quark interactions
\cite{Pirjol:2008gd}.

Using Eqs.~(\ref{ciscalar},\ref{c2scalar}), the observable coefficients $\tilde c_{1,2}$ for the most general scalar
interaction are given by
\begin{eqnarray}\label{citildescalar}
\tilde c_1 &=& \frac16 (R^{(\nu)}_S + R^{(\nu)}_{MS}) \left\{
\begin{array}{c}
1 \\
-1 \\
-\frac14 + \frac{1}{2F} \\
\end{array}
\right\}_\nu \,,\\
\tilde c_2 &=& \frac16 (R^{(\nu)}_S + R^{(\nu)}_{MS}) \left\{
\begin{array}{c}
-\frac32 \\
\frac32 \\
\frac38 - \frac{3}{4F} \\
\end{array}
\right\}_\nu + \frac16 (R^{(\nu)}_S - R^{(\nu)}_{MS}) \left\{
\begin{array}{c}
\frac32 \\
\frac32 \\
-\frac38 - \frac{3}{4F} \\
\end{array}
\right\}_\nu \,.
\end{eqnarray}
In the limit of a zero range scalar interaction, using the relation Eq.~(\ref{RSMS}), this gives
\begin{eqnarray}
\tilde c_1 &=& c_1 - \frac12 c_3 = 0\,,\\
\tilde c_2 &=& c_2 + c_3 = \frac13 R^{(\nu)}_S \left\{
\begin{array}{c}
\frac32 \\
\frac32 \\
-\frac38 - \frac{3}{4F} \\
\end{array}
\right\}_\nu \,.
\end{eqnarray}

We observe that, regardless of the spin-flavor structure of the scalar operator, the coefficient
$\tilde c_1 = c_1 - \frac12 c_3$ vanishes in the limit of a contact interaction. 

We discuss next the extraction of $\tilde c_{1,2}$ from data and examine how well is this suppression 
for $\tilde c_1$ satisfied.
 
\begin{table}
\begin{tabular}{ccccccc} 
\hline
\hline  
$N_{1/2}(1535)$ & $N_{1/2}(1650)$ & $N_{3/2}(1520)$ & $N_{3/2}(1700)$ & $N_{5/2}(1675)$&$\Delta_{1/2}(1620)$ & $\Delta_{3/2}(1700)$ \\ 
\hline
$1535\pm 10$ & $1658\pm 13$ & $1520\pm 5$ & $1700\pm 50$ & $1675 \pm 5$ & $1630\pm 30$ & $1710\pm 40$ \\
\hline 
\hline
\end{tabular} 
\caption{The experimental masses (in MeV) of the $L=1$ non-strange excited baryons from Ref.~\cite{Amsler:2008zzb}. } 
\label{TableII} 
\end{table}

One first estimate can be made using the mixing angles $\theta_{N1, N3}$ determined from a fit to
$N^*$ strong decays and photoproduction data 
$(\theta_{N1}, \theta_{N3}) = (0.39 \pm 0.11, 2.82 \pm 0.11) = (22^\circ \pm 6^\circ,
162^\circ \pm 6^\circ)$ \cite{Goity:2004ss,Scoccola:2007sn}.
Substituting these values into Eqs.~(\ref{c1tildeexp}), (\ref{c2tildeexp}), and using the 
hadron masses from the PDG \cite{Amsler:2008zzb} given in Table~\ref{TableII} we obtain
\begin{eqnarray}\label{exp}
\tilde c_1 &=& 3.9 \pm 11.0 \mbox{ MeV} \,, \\
\tilde c_2 &=& 129 \pm 18 \mbox{ MeV}\,.\nonumber
\end{eqnarray}
This shows that indeed $\tilde c_1$ is suppressed relative to $\tilde c_2$.

\begin{figure}[t] 
\psfrag{A}[][][1.5]{$\tilde c_1$}
\psfrag{B}[][][1.5]{$\tilde c_2$}
\psfrag{T}[t][][1.5]{$\theta_{N1}$}
   \centering
$
\begin{array}{c}
 \includegraphics*[width=10cm]{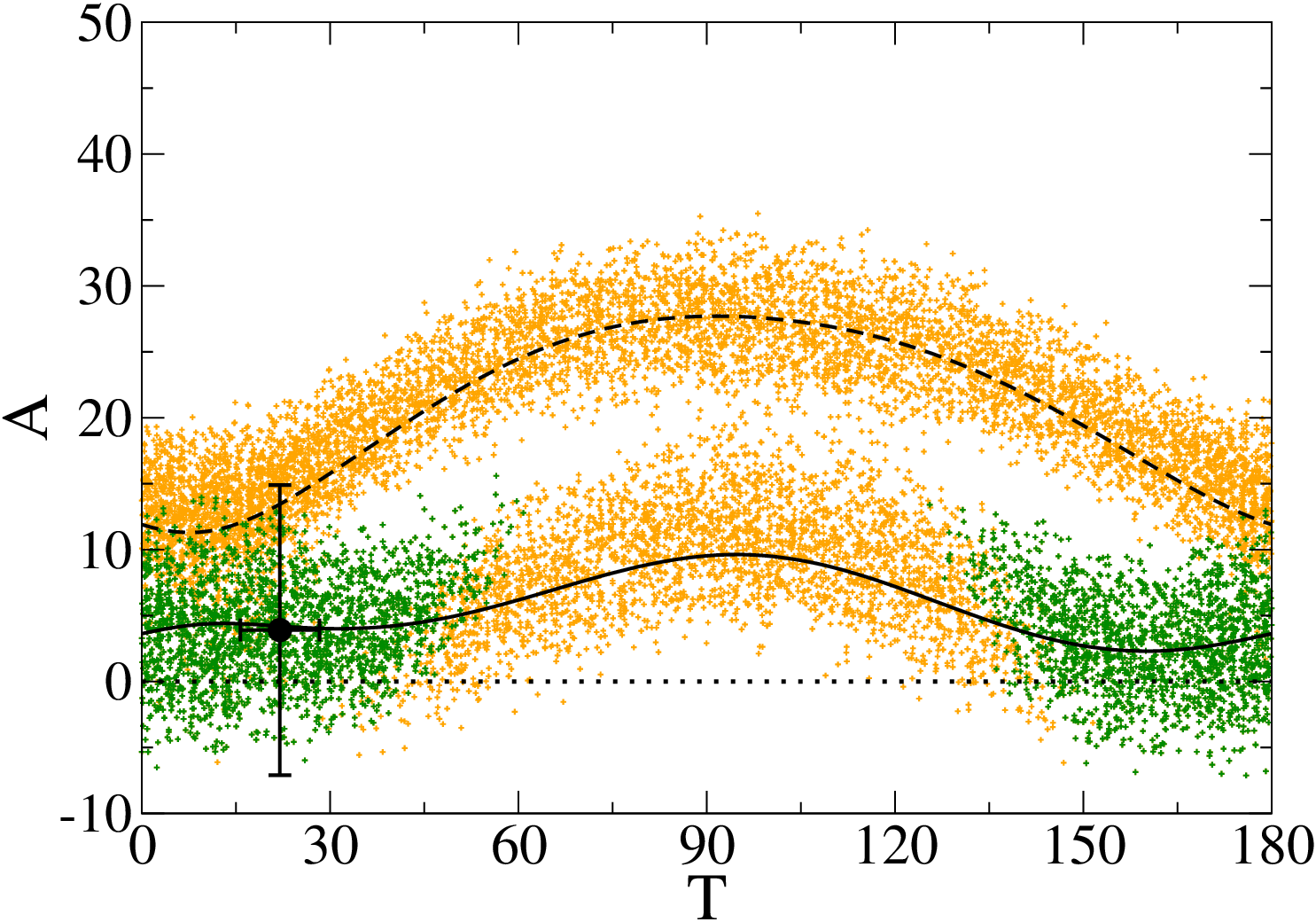} \\
\\ 
 \includegraphics*[width=10cm]{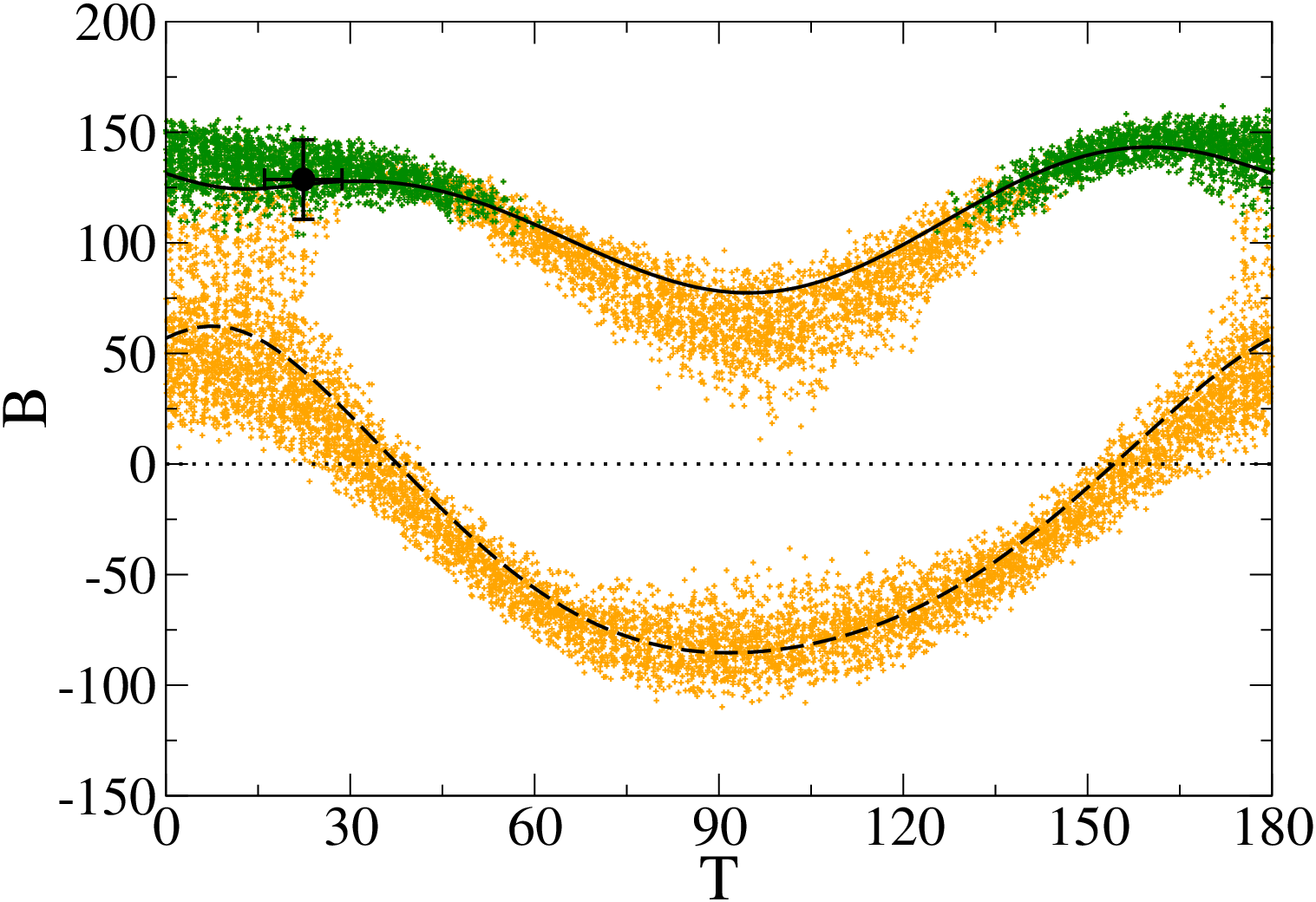} 
\end{array}
$
\caption{The coefficients $\tilde c_{1,2}$ (in MeV)  as a function of the mixing angle $\theta_{N1}$ 
as given by Eqs.~(\ref{c1tildeexp}), (\ref{c2tildeexp}), (\ref{correlation}).
The black points with error bars show the values in Eq.~(\ref{exp}). The dark points (green) of the scatter 
plots give the values allowed by 
imposing the $\bar \Lambda$ constraint as explained in the text. 
The solid and dashed lines correspond to the central values of the masses.
}
   \label{fig:c1t}
\end{figure}

An alternative determination of these coefficients can be made using only the excited baryon masses, 
as discussed in Ref.~\cite{Pirjol:2008gd}. In this paper it was shown that in 
any quark model containing only two-body quark interactions, the mixing angles are correlated (up to a discrete ambiguity) 
by Eq.~(\ref{correlation}) and by a second  relation (Eq.~(6) in Ref.~\cite{Pirjol:2008gd}) 
expressing the spin-average of the $SU(3)$ singlet states $\bar\Lambda = \frac13 \Lambda_{1/2} + \frac23 \Lambda_{3/2}$ 
in terms of the nonstrange states. Allowing for a conservative $SU(3)$ breaking correction of $100 \pm 30$ MeV
in the relation for $\bar\Lambda$, we show in Fig.~\ref{fig:c1t} the scatter plots for $\tilde c_{1,2}$ which 
impose the correlation Eq.~(\ref{correlation}) (all points), and also the relation for $\bar\Lambda$, satisfied on
the dark shaded area (green points). The preferred solution is given by the solid line, which overlaps with the 
dark shaded area (green points). 

We note that there is good agreement between the allowed values of the coefficients $\tilde c_{1,2}$ in the scatter
plots and the results in Eq.~(\ref{exp}), which are shown as the black point with error bars 
on the plot. Both these computations confirm the
suppression of the coefficient $\tilde c_1$ relative to its natural size. The nonvanishing of $\tilde c_1$
can be related to a smearing out of the contact interaction. This is examined in the next Section, where 
it is also found that the sign of the ratio $\tilde c_1 / \tilde c_2$ can provide information on the 
spin-flavor structure of the interaction.

\begin{figure}[b] 
\psfrag{XXX}[][][1.5]{$\tilde c_1/\tilde c_2$}
\psfrag{T}[t][][1.5]{$\theta_{N1}$}
   \centering
  \includegraphics*[width=10cm]{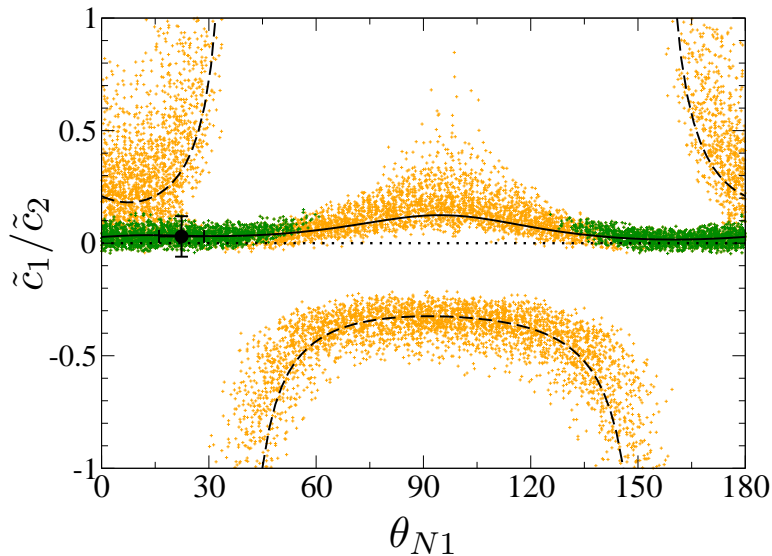} 
   \caption{Scatter plot for the ratio of
coefficients $\tilde c_1/\tilde c_2$ as a function of the mixing angle $\theta_{N1}$. 
The dark (green) points are favored by all data on the hadronic
masses and overlap with the solid line that corresponds to the  
 preferred solution of the correlation Eq.~(\ref{correlation}). The black point with error bars shows  the
values of the coefficients quoted in Eq.~(\ref{cratio}).}
   \label{fig:ratioc12}
\end{figure}

\section{Finite range scalar quark interactions}
\label{finiter}

In the general case of a finite range interaction, the orbital reduced matrix elements $R_S$ and
$R_{MS}$ are independent. According to Eq.~(\ref{citildescalar}) the coefficient $\tilde c_1$ will be in
general
nonvanishing, and proportional to the combination of the reduced matrix elements $R_S + R_{MS}$.
The ratio of $\tilde c_1/\tilde c_2$ for the different spin-flavor structures can be expressed in 
terms of the ratio of reduced matrix elements
\begin{eqnarray}
r_\nu \equiv \frac{R_S^{(\nu)} + R_{MS}^{(\nu)}}{R_S^{(\nu)} - R_{MS}^{(\nu)}}\,.
\end{eqnarray}

We start by considering the case when the quark Hamiltonian contains a single spin-flavor structure 
$O^{(\nu)}_{ij}$. For the pure OGE interaction $O_{ij}^{(2)} = \vec s_i \cdot \vec s_j$  the ratio of coefficients 
$\tilde c_1/\tilde c_2$ is predicted to be
\begin{eqnarray}\label{rOGEmu}
\frac{\tilde c_1}{\tilde c_2} |_{s\cdot s} = -\frac23 \frac{r_2}{1+r_2}
\, .
\end{eqnarray}

For the GBE interaction $O^{(3)}_{ij} = \vec s_i \cdot \vec s_j t_i^a t_j^a$ with two light quark flavors $F=2$, the coefficient $\tilde c_1$ is predicted
to be exactly zero, independently of the spatial form of the  quark interaction. With three light quark flavors
$F=3$, we have
\begin{eqnarray}
\frac{\tilde c_1}{\tilde c_2} |_{st\cdot st (F=3)} = \frac23 \frac{r_3}{5-r_3}
\, . 
\end{eqnarray}

Finally, for the isospin interaction $O_{ij}^{(1)} = t_i^a t_j^a$, the ratio $\tilde c_1/\tilde c_2$ is 
\begin{eqnarray}
\frac{\tilde c_1}{\tilde c_2} |_{t\cdot t} = \frac23 \frac{r_1}{1-r_1}
\, .
\end{eqnarray}

We will illustrate the effect of a finite range quark interaction by taking the spatial dependence 
of the interaction to be
\begin{eqnarray}\label{fmu}
f(\vec r_{ij}) = A\Big(\delta^{(3)}( \vec r_{ij}) - \mu^2 \frac{e^{-\mu r_{ij}}}{4\pi r_{ij}}\Big) \,, 
\end{eqnarray}
with $A$ a coupling constant.
Such an orbital dependence is generated by the exchange of a meson of mass $\mu$
\cite{Glozman:1995fu}, see Ref.~\cite{Eiglsperger:2007ay} for a detailed derivation.

Adopting the functional form Eq.~(\ref{fmu}), we will assume that  the contribution of the
second term of $O(\mu^2)$ to $R_S - R_{MS}$ is always smaller than that of the first term. 
This is always satisfied if  
$|\phi(\vec r)|^2 \leq |\phi(0)|^2$, where $\phi(\vec r)$ is defined by the squared
wave function integrated over one of its arguments  $|\phi(\vec r_{12})|^2 \equiv
\int dr_{13} |\Phi(\vec r_{12}, \vec r_{13})|^2$.
Under this assumption, the contribution of the second term
to any reduced matrix element is given by
\begin{eqnarray}
\mu^2\int d^3 r  |\phi(\vec r)|^2 \frac{e^{-\mu r}}{4\pi r} \leq
\mu^2 |\phi(0)|^2 \int d^3 r   \frac{e^{-\mu r}}{4\pi r} = |\phi(0)|^2
\end{eqnarray}
and is thus smaller than the contribution of the first term. (In this example the quark interaction
was taken between the quarks $1,2$.) This proves that $R_S-R_{MS}$ is always positive.
Taking into account that the contribution to  $R_S+R_{MS}$  of the first term in Eq.~(\ref{fmu}) vanishes,
it is easy to see that $R_S+R_{MS}$ is negative and therefore the ratio of reduced matrix elements
$r=(R_S + R_{MS})/(R_S- R_{MS})$ is negative. 

The information about the sign of the ratio of reduced matrix elements $r_\nu<0$ is sufficiently predictive 
to distinguish between the models considered above, through the sign of the ratio of the coefficients $\tilde c_1/\tilde c_2$, as shown in 
Table~\ref{TableIII}.
We denoted here with OGE$_\mu$ the vector meson exchange model corresponding to a vector
meson or a constituent gluon with mass $\mu$; the limit $\mu=0$ corresponds to the usual one-gluon exchange (OGE) model.

\begin{table}
\begin{center}
\begin{tabular}{cc}
\hline
\hline
Model & sgn$(\tilde c_1/\tilde c_2)$ \\
\hline
OGE$_\mu$ & $+$ \\
OGE & 0 \\
GBE $(F=2)$ & 0 \\
GBE $(F=3)$ & $-$ \\
$O_{ij} = t^a_i t^a_j$ &  $-$ \\
\hline
\hline
\end{tabular}
\end{center}
\label{TableIII}
\caption{The sign of the ratio of coefficients $\tilde c_1/\tilde c_2$ as a test for the spin-flavor structure of 
the scalar quark interaction. }
\end{table}

The natural size of the ratio $\tilde c_1/\tilde c_2$ is of order ${\cal O}(N_c^0)$. Its very small value 
(see solid line in Fig.~\ref{fig:ratioc12}) cannot 
be explained by power counting in $1/N_c$ and must have a dynamical 
origin. We find that it is suppressed for pion exchange interactions in general
(chiral limit or physical pion mass), as $\tilde c_1=0$ from the spin-flavor 
structure alone. In the case of gluon exchange interactions, its smallness
is related to the spatial extent of the interaction (and not related to 
its spin-flavor structure as in the previous case). In the case of a 
contact spin-spin interaction $\tilde c_1=0$ and the ratio vanishes, but otherwise 
this ratio is different from zero.

We comment on the argument presented in Ref.~\cite{Carlson:1998vx} for the dominance of the
operator $S_c^2$ in the mass operator, and compare it with our conclusions. 
As mentioned in the Introduction, in this paper it was argued that the dominance of the $S_c^2$ 
operator follows by assuming one pion exchange.
This follows from the observation that one particular linear combination of operators is 
equivalent to the unit operator (taking its matrix element on the nonstrange states), up to corrections of $O(1/N_c)$
\begin{eqnarray}
T^2 - S_c^2 + 2 \vec s_1 \cdot \vec S_c = - \frac14 \mathbf{1} + O(1/N_c) \,.
\end{eqnarray}
This identity allows one to eliminate one of the three scalar operators. Choosing to eliminate
$O_3 = \vec s_1 \cdot \vec S_c$, the scalar part of the mass operator reads
\begin{eqnarray}
M = c_0 \mathbf{1} + c_1 O_1 + c_2 O_2 + c_3 O_3 =  c'_0 \mathbf{1} + (c_1 - \frac12 c_3) T^2 + (c_2 + \frac12 c_3) S_c^2 + O(1/N_c^2) \,.
\end{eqnarray}
For the pion exchange interaction (both contact and finite range) we find $c_1 - \frac12 c_3 = 0$,
which confirms the result of Ref.~\cite{Carlson:1998vx} of dominance of $S_c^2$ in the large $N_c$
limit. In our approach, at $N_c=3$,  the dominance of $S_c^2$ is exact 
for pion exchange or any contact interaction. 

Using the numerical values of the coefficients $\tilde c_i$ from Eq.~(\ref{exp}), the value of the 
ratio is 
\begin{eqnarray}\label{cratio}
\tilde c_1/\tilde c_2 = 0.03 \pm 0.09\,. 
\end{eqnarray}
An alternative determination using only hadron
masses is shown in Fig.~\ref{fig:ratioc12}. The solid line is the preferred solution \cite{Pirjol:2008gd} 
and gives a range of values compatible with the first determination
(shown in Fig.~\ref{fig:ratioc12} as the black point with error bars).
The central value is positive and clearly suppressed with respect to its natural size ${\cal O}(N_c^0)$
for any value of the mixing angle $\theta_{N1}$. Its sign favors
a pure vector boson exchange model OGE$_\mu$ with a non-vanishing vector meson mass
$\mu$. However, within the errors, negative values or a vanishing ratio are also allowed,
such that it is difficult to draw a clear conclusion. A more precise determination of the 
mixing angles and hadron masses may sharpen this determination and allow one to 
fix the sign of the ratio. 

We comment briefly on the massive vector boson exchange model OGE$_\mu$, 
which produces a positive ratio $\tilde c_1/\tilde c_2$.
This corresponds to a massive gluon model, previously considered in the literature in 
Refs.~\cite{Cornwall:1981zr,lattice,Alkofer:2000wg,Mathieu:2008me,Cornwall:1982zn}. 
In these works it has been suggested that, in the low energy
limit, an effective gluon mass can be generated by nonperturbative QCD effects.  
In principle an effective gluon mass can be observed through its effect on the
low energy limit of quark forces in the constituent quark model. 
In the next Section we perform a crude model calculation to give an estimate of the 
range of allowed values for the effective gluon mass $\mu$. 

The analysis presented above was limited to the spin-flavor structure of the scalar quark interaction.  
In the spin-orbit sector, it has
been pointed out in Ref.~\cite{Pirjol:2008gd} that the flavor dependent interactions $(s_i \pm s_j) t_i^a t_j^a$ are needed in 
order to reproduce the observed mass spectrum. Also in the tensor sector, flavor dependent
operators are needed \cite{Schat:2001xr, Goity:2002pu} in order to produce a non-zero coefficient of the operator 
$\sim \frac{1}{N_c} L_2^{ij} g^{ia} G_c^{ja}$.
Allowing for a mixture of all possible interactions $\nu=1,2,3$ with strengths $A_\nu$ 
\begin{eqnarray}
H = \sum_{\nu=1}^3 A_\nu \sum_{i<j} 
\Big(\delta^{(3)}(\vec r_{ij}) - \mu_\nu^2 \frac{e^{-\mu_\nu r_{ij}}}{4\pi r_{ij}}\Big) O_{ij}^{(\nu)} \,,
\end{eqnarray}
we get the following general results for  the operator coefficients
\begin{eqnarray}
\tilde c_1 &=& A_1 r_1 - A_2 r_2 + (-\frac14 + \frac{1}{2F}) A_3 r_3 \,,  \\
\tilde c_2 &=& \frac32 A_1 (1 - r_1 ) + \frac32 A_2(1+r_2) + A_3 \left[-\frac38 - \frac{3}{4F} + (\frac38-\frac{3}{4F}) r_3\right] \,.
\end{eqnarray}
No simple conclusions about the relative contributions of the different spin-flavor interactions 
can be drawn in the most general case. For example, assuming a mixture of the OGE and GBE $(F=2)$ interactions
($A_1 = 0$), it is possible to arrange positive values for $\tilde c_{1,2}$ by taking $A_{2,3} > 0$ and $A_2$ 
sufficiently large relative to $A_3$ that the second term in $\tilde c_{1,2}$ dominates over the third one.
It is interesting to notice that in the case of the simultaneous presence of a massless one gluon exchange 
interaction ($r_2=0$) and a finite range one pion exchange interaction the coefficient $\tilde c_1$ vanishes 
independently of their relative strengths.

\section{Isgur-Karl model calculation}
\label{IKcalc}

The finite range effects can be taken into account in a quantitative way by adopting a specific
choice for the hadronic model. For illustration we consider the Isgur-Karl (IK) model \cite{Isgur:1977ef}, 
 which has been
widely used for describing the properties of the excited baryons. The matching of this 
model to the effective operator expansion has also been discussed in detail recently in Ref.~\cite{Galeta:2009pn}.

The IK model describes three constituent quarks interacting by harmonic oscillator
potentials 
\begin{eqnarray}\label{HIK}
H_{0} = \frac{1}{2m} \sum_i p_i^2 + \frac{K}{2} \sum_{i<j} r^2_{ij} \,. 
\end{eqnarray}
This Hamiltonian can be solved by introducing the reduced coordinates
\begin{eqnarray}
\vec \rho = \frac{1}{\sqrt2} (\vec r_1 - \vec r_2)\,,\qquad
\vec \lambda = \frac{1}{\sqrt6} (\vec r_1 + \vec r_2 - 2 \vec r_3) \,.
\end{eqnarray}
Expressed in terms of $\lambda, \rho$, the Hamiltonian Eq.~(\ref{HIK}) has the form
of two independent 3-dimensional oscillators
\begin{eqnarray}
H_0 = \frac{p_\rho^2}{2m} + \frac{p_\lambda^2}{2m} + \frac32 K\rho^2 + \frac32 K\lambda^2 \,.
\end{eqnarray}

The eigenstates with orbital angular momentum $L=1,m=+1$ are 
\begin{eqnarray}
\Psi_{m=+1}^{\rho} &=& -(\rho_1 + i\rho_2) \frac{\alpha^4}{\pi^{3/2}}
\exp\Big( - \frac12 \alpha^2 (\rho^2 + \lambda^2) \Big) \,, \\
\Psi_{m=+1}^{\lambda} &=& -(\lambda_1 + i\lambda_2) \frac{\alpha^4}{\pi^{3/2}}
\exp\Big( - \frac12 \alpha^2 (\rho^2 + \lambda^2) \Big) \,.
\end{eqnarray}
where $\alpha = (3Km)^{1/4}$.

The parameters of the model are \cite{KR}
\begin{eqnarray}\label{IKparams}
m=m_u=m_d=420   \mbox{ MeV} \,,\qquad
\alpha = 467 \mbox{ MeV} \,,\qquad
\alpha_{s}=0.95 \, .
\end{eqnarray}

The reduced matrix elements $R_S, R_{MS}$ are
given by the matrix elements
\begin{eqnarray}
\langle \Psi^\lambda | f(\vec r_{12}) | \Psi^\lambda \rangle = \frac13 (R_S - R_{MS}) \,, \\
\label{Psirho}
\langle \Psi^\rho | f(\vec r_{12}) | \Psi^\rho \rangle = \frac13 (R_S + R_{MS}) \,.
\end{eqnarray}

The combination of reduced matrix elements $R_S - R_{MS}$ was computed in 
Ref.~\cite{Galeta:2009pn} (see Eq.~(52)) in the limit of a contact interaction $\mu=0$. 
Using the spatial dependence of $f(\vec r_{12})$ given in Eq.~(\ref{fmu}) one finds the complete result for $\mu\neq 0$
\begin{eqnarray}
R_S - R_{MS} = A\frac{3\alpha^3}{(2\pi)^{3/2}} \Phi_-(\frac{\mu}{\alpha}) \,, 
\end{eqnarray}
where the function $\Phi_-(x)$ is given by
\begin{eqnarray}
\Phi_-(x) = 1 - x^2 + \sqrt{2\pi} x^3 N(-x) e^{\frac12 x^2}
\end{eqnarray}
and is positive for $x>0$, which confirms that the contribution of the 
finite range term in $\langle \Psi^\lambda |f(\vec r_{12})|\Psi^\lambda\rangle$ is never larger than that of 
the contact term.

The function $N(x)$ is the cumulative normal distribution function, which is related to the
erf$(x)$ function, and is defined as
\begin{eqnarray}
N(x) = \frac12 \left[ 1+\mbox{erf}(\frac{x}{\sqrt2}) \right]  = \int_{-\infty}^x \frac{dy}{\sqrt{2\pi}} e^{-y^2/2} \,.
\end{eqnarray}

In a similar way one can compute also the
combination of reduced matrix elements $R_S + R_{MS}$ which vanishes in the limit of a contact interaction. We obtain
\begin{eqnarray}
R_S + R_{MS} = -A\frac{2\alpha\mu^2}{(2\pi)^{3/2}} \Phi_+(\frac{\mu}{\alpha}) \,,
\end{eqnarray}
where the function $\Phi_+(x)$ is given by
\begin{eqnarray}
\Phi_+(x) = 1 + \frac12 x^2 - \sqrt{\frac{\pi}{2}} x N(-x) e^{\frac12 x^2}(3+x^2)
\end{eqnarray}
and is positive for $x>0$. 

The asymptotic behavior of the functions $\Phi_\pm(x)$ at small and large values of the argument $x$
is
\begin{eqnarray}
\Phi_+(x) &=& 1 - \frac32 \sqrt{\frac{\pi}{2}} x + 2x^2 + O(x^3)\,, \qquad\quad x \ll 1 \\
\Phi_-(x) &=& 1 - x^2 + \sqrt{\frac{\pi}{2}} x^3 + O(x^4)\,, \qquad \qquad x \ll 1 \\
\Phi_+(x) &=& \frac{3}{x^4} - \frac{30}{x^6}+ O(x^{-8})\,, \qquad\qquad\qquad\,\,\,  x \gg 1 \\
\Phi_-(x) &=&  \frac{3}{x^2} - \frac{15}{x^4} + O(x^{-6})\,, \qquad\qquad\qquad\,\,\, x \gg 1 \,.
\end{eqnarray}

\begin{figure}[b!]
\psfrag{r}[][][1.5]{$r$}
\psfrag{x}[t][][1.5]{$\mu/\alpha$}
\includegraphics*[width=10cm]{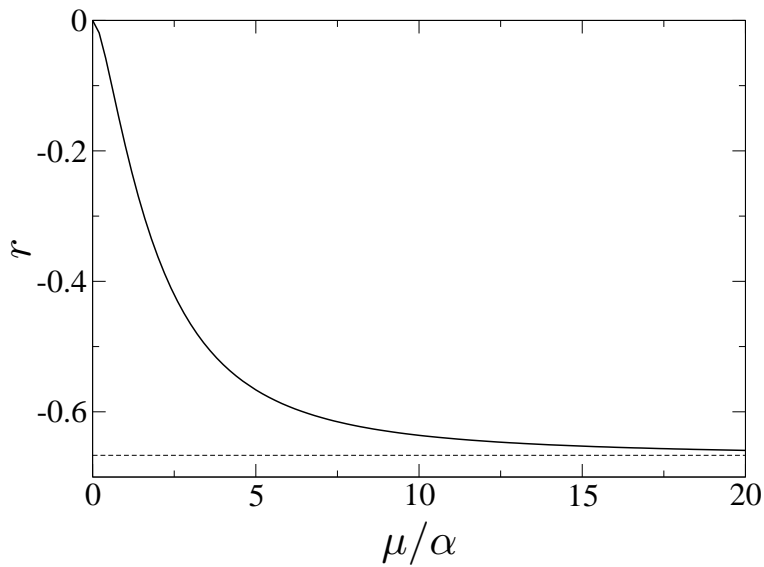}
\caption{\label{fig:r}
The ratio of reduced matrix elements $r=(R_S + R_{MS})/(R_S - R_{MS})$ as a function of the
ratio $\mu/\alpha$ in the IK model, as given by Eq.~(\ref{rIK}).}
\end{figure}

The unknown constant $A$ cancels out in the ratio of reduced matrix elements $r$, 
which depends only on the ratio $\mu/\alpha$
\begin{eqnarray}\label{rIK}
r = \frac{R_S + R_{MS}}{R_S - R_{MS}} = -\frac23 \left( \frac{\mu}{\alpha}\right)^2 
\frac{\Phi_+(\mu/\alpha)}{\Phi_-(\mu/\alpha)} \,.
\end{eqnarray}
We show in Fig.~\ref{fig:r} a plot of the ratio $r$ as a function of $\mu/\alpha$.
The ratio $r$ vanishes in the limit $x = 0$ of a massless exchanged particle, while for an infinitely
heavy mass it approaches a finite limit $r_\infty = -\frac23$. 

Using these results we can obtain constraints on the mass $\mu$ of the exchanged boson. 
We quote results separately for the negative and positive ranges of the ratio $\tilde c_1/\tilde c_2$,
corresponding to the $t_i^a t_j^a$ and OGE$_\mu$ models \footnote{For a negative ratio we also have the GBE (F=3) interaction,
but we don't consider it in the phenomenological discussion as flavor $SU(3)$ is broken.}, respectively.
From Eq.~(\ref{exp}) one finds 
\begin{eqnarray}
\frac{\tilde c_1}{\tilde c_2} &=&
\left\{
\begin{array}{cc}
\ [ -0.06 ,  0.00 ] \,,  & \qquad r_1 = [ -0.10 ,  0.00 ] \,, \\
\ [ \;\;  0.00 , 0.12 ] \,,  & \qquad  r_2 = [ -0.15 ,  0.00 ] \,.
\end{array}
\right.
\end{eqnarray}
Using Eq.~(\ref{rIK}) this can be translated into ranges of allowed values for the boson mass $\mu$, namely
\begin{eqnarray}
t_i^a t_j^a &:& \qquad 0.0 \leq \frac{\mu}{\alpha} \leq 0.58 \nonumber \,,  \\
\mbox{OGE}_\mu &:& \qquad 0.0 \leq \frac{\mu}{\alpha} \leq 0.82 \,.
\end{eqnarray}

Using for the mass scale $\alpha$ the typical value of the Isgur-Karl model given
in Eq.~(\ref{IKparams}), we obtain for the mass of the vector boson which can reproduce the
observed data the allowed range $\mu = [0,383]$ MeV. This is much smaller than the lowest bound for 
a constituent gluon mass  $m_g \simeq 800$ MeV suggested by lattice calculations of hybrid meson masses
\cite{Bernard:2003jd}
and the glueball spectrum \cite{Chen:2005mg}.
The use of the Isgur-Karl model and its parameters is a very crude first attempt 
to give an estimate of $\mu$ in the OGE$_\mu$ case. 
It would be worthwhile to improve on this to see if it is possible to obtain a better 
estimate of $\mu$ that is compatible with the bounds obtained from lattice calculations,
as well 
as an interpretation of the other possible spin-flavor interactions as the result of
quark exchange or meson exchange interactions.

\section{Summary and conclusions}
\label{concl}

The hierarchy of the observed coefficients in the $1/N_c$ expansion for the $L=1$ excited baryons
has a very specific pattern, with one of the subleading $O(1/N_c)$ operator $S_c^2$ dominating, 
and the coefficients of the other operators suppressed. In this paper we present a possible explanation for 
the dominance of the $S_c^2$ operator in the framework of the constituent quark model. 

Considering the most general two-body quark interaction, we show that a contact
quark interaction leads to the suppression of certain coefficients in the $1/N_c$ operator expansion.
Furthermore, any spin-flavor zero range two-body quark interaction is matched onto the same operator
$S_c^2$. 
Intuitively, this can be understood from the fact that the excited and core quarks are in
a relative p-wave, and thus the coefficients of $t^a T_c^a$ and $\vec s\cdot \vec S_c$ vanish if the
spatial part of the interaction is a $\delta^{(3)}(\vec r)$ function \cite{Schat:2001xr}.
This result implies that it is impossible to distinguish between different models of quark 
interactions as long as they are of zero range.

Allowing for a quark interaction of finite range, modeled by the exchange of a particle of
mass $\mu$, we study the question of obtaining information about the spin-flavor structure
of the scalar part of the quark interaction from the mass spectrum of the negative
parity $L=1$ excited baryons. Under the assumption that only one spin-flavor structure dominates,
we find that the sign of the ratio $\tilde c_1/\tilde c_2$, that can be obtained from the 
experimental masses and mixing angles, can be used as a test of the spin-flavor structure of the 
interaction. 

The central value we obtain for this ratio corresponds to a spin-spin interaction with the exchange of a massive vector
boson. Using the wave functions of the Isgur-Karl model, the mass of the exchanged vector
meson is in the range $\mu \sim [0,400]$ MeV, which is much smaller than the lowest bound 
for the mass of a constituent gluon, as suggested by lattice calculations of hybrid 
meson masses \cite{Bernard:2003jd} and glueballs \cite{Chen:2005mg}. This seems to disfavor 
this type of interaction. 
The present study shows that within
error bars and allowing for two scalar spin-flavor structures, the one (massless) gluon exchange and the one pion 
exchange interactions lead to $\tilde c_1=0 $ independently of their relative strengths, 
and are consistent with data. Previous studies 
\cite{Schat:2001xr,Goity:2002pu,Pirjol:2008gd}
that did not focus on the range of the microscopic interaction also include the spin-orbit and tensor terms, 
and suggest that both gluon exchange and flavor dependent interactions are 
needed to reproduce the data.  
Allowing for a more general combination of spin-flavor structures and finite range forces,
no useful information on their contribution is obtained from the present analysis in the absence of additional
dynamical information about their relative strength.

\begin{acknowledgments}
The work of C.S. was supported by CONICET and partially supported by  the U.~S.  Department of Energy, Office of
Nuclear Physics under contract No. DE-FG02-93ER40756
with Ohio University.
\end{acknowledgments}

\end{document}